\documentclass[AMA,STIX1COL]{WileyNJD-v2}
\newcommand\numberthis{\addtocounter{equation}{1}\tag{\theequation}}
\colorlet{BLUE}{blue}
\colorlet{GRAY}{gray}
\usepackage{lscape}
\usepackage{pdflscape}
\articletype{Research Article}%

\received{}
\revised{}
\accepted{}

\begin{document}
\title{Use of copula to model within-study association in bivariate meta-analysis of binomial data at the aggregate level: a Bayesian approach and application to surrogate endpoint evaluation.}

\author[1,2]{Tasos Papanikos}
\author[3]{John R Thompson}
\author[1,4]{Keith R Abrams}
\author[1]{Sylwia Bujkiewicz}

\authormark{Papanikos \textsc{et al}}

\address[1]{Biostatistics Group, Department of Health Sciences, University of Leicester, Leicester, UK}
\address[2]{GlaxoSmithKline R\&D centre, GlaxoSmithKline, Stevenage, UK}

\address[3]{Genetic Epidemiology Group, Department of Health Sciences, University of Leicester, Leicester, UK}
\address[4]{Department of Statistics, University of Warwick, , Coventry, UK}

\corres{Tasos Papanikos,  \email{anastasios.x.papanikos@gsk.com}}

\presentaddress{Biostatistics Research Group \\ Department of Health Sciences \\ University of Leicester \\ George Davies Centre \\ University Road \\ Leicester \\ LE1 7RH \\ United Kingdom}

\abstract[Abstract]{	
Bivariate meta-analysis provides a useful framework for combining information across related studies and has been utilised to combine evidence from clinical studies to evaluate treatment efficacy on two outcomes. It has also been used to investigate surrogacy patterns between treatment effects on the surrogate endpoint and the final outcome. Surrogate endpoints play an important role in drug development when they can be used to measure treatment effect early compared to the final outcome and to predict clinical benefit or harm. The standard bivariate meta-analytic approach models the observed treatment effects on the surrogate and the final outcome outcomes jointly, at both the within-study and between-studies levels, using a bivariate normal distribution. For binomial data, a normal approximation on log odds ratio scale can be used. However, this method may lead to biased results when the proportions of events are close to one or zero, affecting the validation of surrogate endpoints. In this paper, we explore modelling the two outcomes on the original binomial scale. Firstly, we present a method that uses independent binomial likelihoods to model the within-study variability avoiding to approximate the observed treatment effects. However, the method ignores the within-study association. To overcome this issue, we propose a method using a bivariate copula with binomial marginals, which allows the model to account for the within-study association.  We applied the methods to an illustrative example in chronic myeloid leukemia to investigate the surrogate relationship between complete cytogenetic response  (CCyR) and event-free-survival (EFS).
}

\maketitle

	\section{Introduction}\label{intro}

	Bivariate meta-analytic methods provide a natural framework for synthesising evidence obtained from two outcomes. When meta-analysing correlated outcomes, two sources of association exist in the data, one at the individual level and one at the study level. Specifically, within each study, the treatment effects on the two outcomes are measured on the same individuals and therefore are correlated (within-study correlation). Additionally, the between-studies variability on the first and the second outcome (due to, for example, differences in study population or treatment dose) generate correlation at the between-studies level (between-studies correlation) \cite{riley2017multivariate}.	
	
	A bivariate random effects meta-analysis model (BRMA) \cite{van2005tutorial} can be used to perform bivariate meta-analysis of correlated and normally distributed treatment effects on two outcomes. This method models treatment effects on both outcomes jointly with a bivariate normal distribution. A very popular form of the bivariate normal meta-analytic method has been described by Houwelingen et al. \cite{van2005tutorial} and Riley et al.\cite{riley2007evaluation}. This approach accounts for the within-study correlation and it can be used to obtain mean treatment effects on both outcomes, as well as, to assess the study-level association between the treatment effects between the first and the second outcome \cite{BujkiewiczTSD,bujkiewicz2015uncertainty}. When this approach is applied to binomial data, the proportions of events in each arm across outcomes can be transformed to obtain treatment effects on log odds ratio (OR) scale, which are assumed to be approximately normally distributed.
	However, when modelling binomial data on log OR scale, the assumption of normality may not always be reasonable. Hamza et al. \cite{hamza2008binomial} showed that the normal approximation, used for binomial data in univariate meta-analysis of diagnostic test accuracy studies leads to biased results, especially when the proportions of events are very close to zero or one and the variance is large. A similar issue is likely to occur when modeling binomial responses to treatment.  Whilst synthesis of a single binomial outcome data using exact binomial likelihood is straightforward, a bivariate meta-analysis is challenging, unless some structure of the data is present (such as outcomes are mutually exclusive or have an is-subset-of relationship) and taken into account \cite{trikalinos2014empirical}.
	
	In this paper, we investigate the importance of the choice of the scale and the corresponding distributional assumptions when modelling bivariate binomial data in a meta-analytic framework in the context of surrogate endpoint evaluation. Bivariate meta-analysis of treatment effects on a surrogate endpoint and a final outcome allows for the study level validation of a surrogate endpoint.	
	A standard way to validate the study level surrogacy is to perform a form of bivariate meta-analysis, such as BRMA, to model jointly correlated and normally distributed treatment effects on surrogate and final outcomes \cite{van2005tutorial,daniels1997meta} and monitor the between-studies correlation parameter.  When study level validation of surrogate endpoints is based on data from modern clinical trials assessing personalized treatments, the high effectiveness of such targeted therapies results in large proportions of responders and very small proportions of progressions or deaths. Therefore, the assumption of normality when modeling binomial aggregate data on effectiveness of such therapies may lead to poor inferences about the parameters describing the surrogate relationship and may affect the study level validation of a surrogate endpoint. This may have a significant impact on regulatory decisions about market access of new therapies, in particular when the poor choice of modeling assumptions can lead to over/underestimation of the correlation. 
	
	To address this issue, we present two random effect meta-analytic methods for the evaluation of study level surrogate relationships of the treatment effects on binomial outcomes, using exact likelihood approach based on the binomial distribution when modelling within-study variability. The first approach is an modification of a generalised liner mixed model (GLMM) applied to meta-analysis of diagnostic accuracy studies \cite{chu2006bivariate}, extending the method to model data from comparative studies. It uses the exact independent binomial likelihoods across outcomes and treatment arms to model the within-study variability. This model, however, ignores potential within-study associations. In a previous work, Riley et al. \cite{riley2009multivariate} highlighted the importance of taking into account the within-study correlation when using BRMA model. To account for the within-study association on the binomial scale, we introduce another method which models the summarised events on each outcome jointly using a bivariate copula with binomial marginal distributions. This model accounts for the within-study association between the summarised events on the surrogate and the final outcome through the copula dependence parameter. This makes the copula model a more appropriate approach, compared to one using independent binomial likelihoods, as the events on the surrogate endpoint and the final outcome are obtained from the same patients and therefore, they are correlated. Copulas have been previously used to model individual level surrogacy patterns modelling dependencies between, for example, time to event surrogate and final outcomes in individual patient data (IPD) based methods \cite{burzykowski2004validation}. IPD, however, are often not available, and only study level surrogacy can be validated using summary data. Thus robust methods for the synthesis of aggregate data for surrogate endpoint evaluation are very important. 
	
	We investigate the impact of assumptions made when modelling the within-study variability on the estimates of the between-studies parameters in the meta-analysis of two binomial outcomes (surrogate and final) and in particular when the proportions of events (such as responses to treatment or deaths) are close to zero or one. We carry out this investigation in a simulation study, comparing the two proposed methods and the standard BRMA approach, as well as by applying the methods to an illustrative example in chronic myeloid leukemia (CML). 
	
	The illustrative data example in CML is introduced Section 2. Section 3 discusses existing and the proposed methodology providing also a short overview of the copula theory. Section 4 presents the results of fitting the models to the data of the motivating example. To illustrate the motivation and the application of the proposed method in a more detailed and controlled manner and to compare its performance against the existing models, we carried out a simulation study. The design and the results of the simulation study are discussed in Section 5. The paper concludes with a discussion in Section 6. 
	
	\section{Data example}\label{Example}
	
	CML is a myeloproliferative neoplasm of hematipoietic stem cells associated with a characteristic chromosomal translocation called the Philadelphia chromosome. The main characteristic is that CML is regarded as a slow progressive disease \cite{baccarani2010chronic}. Before the molecular pathogenesis of the disease was well understood, the median survival was 6 years, with a predicted 5-year overall survival (OS) of 47.2\% \cite{huang2012estimations}. However, the introduction of tyrosine kinase inhibitor (TKI) \cite{o2003imatinib} therapies has led to dramatically improved long-term survival rates resulting in high response rates of complete cytogenetic response (CCyR) at 12 months and very few events such as loss of response (e.g complete cytogenetic response, major molecular response etc.), progression to accelerated phase (AP) or blast crisis (BC) and death from any cause. To illustrate alternative modeling approaches and compare them with BRMA, we identified 10 studies comparing first generation TKI therapies (e.g 400mg imatinib) with second generation TKIs (e.g. dasatinib, nilotinib, busotinib) or different doses of first generation TKIs (600mg or 800mg imatinib). We evaluated the study level association between two binary outcomes, CCyR at 12 months and event-free survival (EFS) at 24 months. We chose CCyR at 12 months as it has been extensively used in the literature as a gold standard for a good measure of response and EFS at 24 months as it is very significant in view of the dismal prognosis of the patients proceeding to advanced stages or losing response. 
	Table 3 presents the summarised responses in the treatment and the control arms on both outcomes along with the sample size per arm and outcome. We chose to work with positive correlations and to do so, we modelled the number of patients who were event-free at 24 months EFS.
	
	\begin{table}[h!]
		\centering
		\label{Studies}
		\caption{Summarised data in CML}
		\begin{tabular}{lccccccccc}   
			\hline 	
			&\multicolumn{4}{c}{Complete Cytogenetic Response}&&\multicolumn{4}{c}{Event-Free-Survival}\\
			\cline{2-5}\cline{7-10}
			&\multicolumn{2}{c}{Control arm}&\multicolumn{2}{c}{Treatment arm}&&\multicolumn{2}{c}{Control arm}&\multicolumn{2}{c}{Treatment arm}\\
			\cline{2-3}\cline{4-5}\cline{7-8}\cline{9-10}
			Study name     & Arm size&Responses&Arm size& Responses&&Arm size& Events&Arm size& Events\\	
			\hline
			Cortes 2012 \cite{cortes2011bosutinib}     &252 &171 &250 &175 & &252 &222 &250 &230 \\
			Kantarjian 2010 \cite{kantarjian2012dasatinib} &260 &189 &259 &216 & &260 &239 &259 &243 \\
			Radich 2012 \cite{radich2012randomized}    &61  &42  &70  &59  & &123 &117 &123 &118 \\
			Saglio 2010 \cite{kantarjian2011nilotinib}    &243 &184 &236 &219 & &283 &267 &281 &276 \\
			Baccarani 2009 \cite{baccarani2009comparison} &108 &63  &108 &69  & &108 &74  &108 &77 \\
			Preudhomme 2010 \cite{preudhomme2010imatinib} &158 &92  &160 &104 & &159 &149 &160 &149 \\
			Hehlmann 2011 \cite{hehlmann2011tolerability}   &306 &151 &328 &206 & &324 &308 &338 &317 \\
			Cortes 2010 \cite{cortes2010phase}     &157 &103 &319 &223 & &157 &149 &319 &311 \\
			Deininger 2013 \cite{deininger2014imatinib}  &49  &33  &41  &35  & &73  &68  &72  &60 \\
			Wang 2015 \cite{wang2015phase}      &133 &107 &134 &104 & &133 &125 &134 &124 \\ 
			\hline
		\end{tabular} 
	\end{table}

	\section{Methods}\label{methods}
	\subsection{Bivariate random effects meta-analysis (BRMA)}\label{BRMA}
	The BRMA model for correlated and normally distributed treatment effects on
	two outcomes $Y_{1i}$ and $Y_{2i}$ was firstly introduced by McIntosh \cite{mcintosh1996population} and since then many extensions have been proposed. It is usually presented in the form described by van Houwelingen \emph{et al.} \cite{van2005tutorial} and Riley \emph{et al.} \cite{riley2007evaluation}:
	\begin{gather*}
	\begin{pmatrix}Y_{1i}\\
	Y_{2i}
	\end{pmatrix} \sim  N
	\begin{pmatrix}
	\begin{pmatrix}
	\delta_{1i}\\
	\delta_{2i}
	\end{pmatrix}\!\!,&
	\begin{pmatrix}
	\sigma_{1i}^{2} & \sigma_{1i}\sigma_{2i}\rho_{wi} \numberthis \label{eq:1}\\
	\sigma_{1i}\sigma_{2i}\rho_{wi} & \sigma_{2i}^{2}
	\end{pmatrix}
	\end{pmatrix}\\
	\begin{pmatrix}\delta_{1i}\\
	\delta_{2i}
	\end{pmatrix} \sim  N
	\begin{pmatrix}
	\begin{pmatrix}
	d_{1}\\
	d_{2}
	\end{pmatrix}\!\!,&
	\begin{pmatrix}
	\tau_{1}^{2} & \tau_{1}\tau_{2}\rho_{b}\numberthis \label{eq:2} \\
	\tau_{1}\tau_{2}\rho_{b} & \tau_{2}^{2}
	\end{pmatrix}
	\end{pmatrix}
	\end{gather*}
	
	In this model, the treatment effects on the first and the second outcome $Y_{1i}$, $Y_{2i}$, which can be log OR, are
	assumed to estimate the correlated true treatment effects $\delta_{1i}$ and $\delta_{2i}$ with corresponding within-study variances
	$\sigma_{1i}^2$ and $\sigma_{2i}^2$  of the estimates and the within-study correlation $\rho_{wi}$
	between them.
	In this hierarchical framework, these true study-level effects follow a bivariate normal distribution with means $\left(d_1, d_2\right)$ corresponding to the two outcomes, the between-studies variances $\tau_{1}^{2}$ and $\tau_{2}^{2}$ and the between-studies correlation $\rho_b$. In the context of surrogate endpoints the between-studies correlation $\rho_{b}$ is the main parameter of interest and it is used to assess the study level association between the treatment effect on the surrogate endpoint and the effect on the final outcome.
	Equation (\ref{eq:1}) represents the within-study model and (\ref{eq:2}) is the between-studies model.
	
	The elements of the within-study covariance matrix, $\sigma_{1i}^2$, $\sigma_{2i}^2$ and $\rho_{wi}$  are assumed to be known. Whilst the estimates of the variances are easily obtained by taking the square of the standard error for each outcome, the estimates of the within-study correlations between the treatment effects on the two outcomes are more difficult to obtain as they would not be reported in the original articles. When IPD are available, the correlation can be obtained for normally distributed outcomes by, for example, fitting a regression model for the two outcomes with correlated errors \cite{riley2015multivariate}. For transformed binomial or time to event outcomes (such as log OR or log HR) the within-study correlation can be estimated by bootstrapping (see details in the section \ref{BootMeth} and section A.1 of the supplementary material). Other methods of estimating the within-study correlations have been discussed elsewhere and are summarized in Bujkiewicz et al \cite{BujkiewiczTSD}.
	Implementing the model in the Bayesian framework the unknown parameters $\tau^{2}_{1}$, $\tau^{2}_{2}$, $d_{1}$, $d_{2}$ and $\rho_{b}$ have to be estimated and therefore, prior distributions should be specified for them. For instance, the following prior distributions can be placed on the these parameters: $d_{1,2}\sim N(0,10^{2})$, $\tau_{1,2}\sim U(0,5)$, to implement the natural constrain of $-1\le\rho_{b}\le1$ we used the Fisher's $z$ transformation as, $\rho_b = tanh(z)$ , $z\sim N(0,1)$.

	When this model is applied to binomial aggregate data, the data are transformed to obtain treatment effects on the log OR scale: $Y_{1i} = log(\frac{r_{1Bi}}{N_{Bi}-r_{1Bi}})-log(\frac{r_{1Ai}}{N_{Ai}-r_{1Ai}})$,\  $Y_{2i} = log(\frac{r_{2Bi}}{N_{Bi}-r_{2Bi}})-log(\frac{r_{2Ai}}{N_{Ai}-r_{2Ai}})$ with corresponding the variances: $\sigma_{1i}^{2} = \frac{1}{r_{1Bi}}+\frac{1}{N_{Bi}-r_{1Bi}}+\frac{1}{r_{1Ai}}+\frac{1}{N_{Ai}-r_{1Ai}}$ and $\sigma_{2i}^{2}= \frac{1}{r_{2Bi}}+\frac{1}{N_{Bi}-r_{2Bi}}+\frac{1}{r_{2Ai}}+\frac{1}{N_{Ai}-r_{2Ai}}$, where $r_{1Ai}$, $r_{2Ai}$, $r_{1Bi}$, $r_{2Bi}$ are the numbers of events in the control arm $A$ and treatment arm $B$ on both outcomes in study $i$, whereas $N_{Ai}$ and $N_{Bi}$ are the arm sizes in study $i$. 
	A modelling issue occurs when there are no events in either of the treatment arms as the log odds ratios ($Y_{1i}$, $Y_{1i}$) and their variances cannot be defined. A very simple way to tackle this problem is to apply a correction, for instances, by adding 0.5. However, in some situations the effect of adding 0.5 may lead to biased results \cite{cox1970analysis,moses1993combining}. Furthermore, when the proportions of events are close to zero or one the assumption of normality of log ORs is unreasonable and can lead to biased results \cite{hamza2008binomial}. To address these issues, we explore two alternative approaches to modeling Binomial data using exact Binomial likelihood, which are described in the following two sections; one approach  ignoring the within-study correlation and one applying a copula to account for the association at the within-study level.
	
	\subsection{Bivariate random effect meta-analysis with independent Binomials (BRMA-IB)}\label{IndepModel}
	
	In this section, we present a bivariate meta-analytic model with independent binomial likelihoods for two outcomes at the within-study level. This approach is very similar to a standard generalised linear mixed effects model (GLMM) for meta-analysis of diagnostic test accuracy studies \cite{chu2006bivariate,harbord2006unification} (where true positive and true negative observations are not correlated within a study as they are obtained from different individuals). To adapt GLMM to the context of bivariate meta-analysis of randomised clinical trials, we assume that the numbers of events $r_{1Ai}$, $r_{2Ai}$, in the control arm $A$ and $r_{1Bi}$, $r_{2Bi}$ in the experimental arm B, on the two outcomes (the surrogate and the final outcome respectively) follow independent Binomial distributions with the corresponding true probabilities of events  $p_{1Ai}$, $p_{2Ai}$, $p_{1Bi}$ and $p_{2Bi}$:
	\begin{gather*}
	r_{1Ai}\sim Bin(p_{1Ai},N_{Ai}),\quad r_{2Ai}\sim Bin(p_{2Ai},N_{Ai}),\quad
	r_{1Bi}\sim Bin(p_{1Bi},N_{Bi}),\quad r_{2Bi}\sim Bin(p_{2Bi},N_{Bi})\numberthis\label{eq:3}\\
	\end{gather*}
	At the between-studies level (\ref{eq:4}), the true probabilities of events are transformed using a link function $g(\cdot)$ (e.g. logit). 
	\begin{gather*}
	g(p_{1Ai})=\mu_{1i},\quad g(p_{1Bi})=\mu_{1i}+\delta_{1i}\\
	g(p_{2Ai})=\mu_{2i},\quad g(p_{2Bi})=\mu_{2i}+\delta_{2i}\\
	\begin{pmatrix}\delta_{1i}\\
	\delta_{2i}
	\end{pmatrix} \sim  N
	\begin{pmatrix}
	\begin{pmatrix}
	d_{1}\\
	d_{2}
	\end{pmatrix}\!\!,&
	\begin{pmatrix}
	\tau_{1}^{2} & \tau_{1}\tau_{2}\rho_{b} \numberthis\label{eq:4}\\
	\tau_{1}\tau_{2}\rho_{b} & \tau_{2}^{2}
	\end{pmatrix}
	\end{pmatrix},
	\end{gather*}
	where $\mu_{ji}$ are the study specific baseline effects (i.e. the log-odds for the control group $A$ and outcome $j=1,2$ in study $i$) while, $\delta_{ji}$ are the study specific correlated true treatment effects on the log OR scale for outcome $j=1,2$ in study $i$. $\left(d_{1},d_{2}\right)$ are the mean treatment effects on first and the second outcome, $\tau_{1}$ and $\tau_{2}$ are the between-studies heterogeneity parameters and $\rho_b$ the between-studies correlation. Similarly as in the BRMA, between-studies correlation quantifies the relationship between the surrogate endpoint and the final outcome.
	
	To implement the model in the Bayesian framework, we place prior distributions on unknown parameters including the baseline treatment effects $\mu_{1i,2i}\sim N(0,10^{2})$, the mean effects $d_{1,2}\sim N(0,10^{2})$, the between-studies standard deviations $\tau_{1,2}\sim U(0,5)$ and $\rho_b = tanh(z)$ , $z\sim N(0,1)$.
	
	The main difference between this method and the BRMA model is the within-study level (\ref{eq:3}). BRMA-IB models the within-study variability using the exact likelihood approach based on the binomial distribution avoiding to make the restrictive assumption of normality. This approach does not require continuity corrections, however, the model ignores the within-study association which is restrictive as within each study the treatment effects on the two outcomes are measured on the same individuals and hence are correlated. As discussed above, when modeling aggregate data obtained from correlated binary outcomes, two sources of association exist; one at the individual level and one at the study level, and BRMA-IB model accounts only for the latter.
	
	\subsection{Model with bivariate copula}\label{Cop}
	In this section, we propose a novel method using a copula representation to model the within-study variability in such a way to allow for the association between the numbers of events in each arm on the first and the second outcome to be taken into account. Copulas are flexible tools for modeling multivariate data as they account for the dependencies between multiple outcomes and allow for different dependence structures, avoiding the restrictive assumption of normality. Firstly, we introduce some background on copula models. The new model based on copulas is presented in Section \ref{copmodel}.

	\subsubsection{Overview of copula theory}\label{copulatheory}
	A bivariate copula $C$ is a bivariate cumulative distribution function (CDF) restricted to the unit square with standard uniform marginal distributions \cite{joe1997multivariate, joe2014dependence, nelsen2006introduction}.
	
	If $H$ is a bivariate CDF with univariate CDF margins $F_{1}$, $F_{2}$  then according to the Sklar's theorem \cite{sklar1959fonctions} for every bivariate distribution there exists a copula representation $C$ such that
	\begin{equation}
	\label{copuladef1}
	H(x_{1},x_{2},\theta) = C(F_{1}(x_{1}),F_{2}(x_{2}),\theta)
	\end{equation}
	The copula $C$ is unique if $F_{1}$, $F_{2}$ are continuous random variables; otherwise, there are many possible copulas if some of the margins have discrete components as emphasized by Genest and Neslehova \cite{genest2007primer} but all coincide on the closure of $Ran(F_{1})\times Ran(F_{2})$ where $Ran(F)$ denotes the range of $F$. The discrete bivariate probability mass function (pmf) can written in the following form:
	
	\begin{align}
	\label{copula}
	h(x_{1},x_{2},\theta) &= C(F_{1}(x_{1}),F_{2}(x_{2}),\theta)-C(F_{1}(x_{1}-1),F_{2}(x_{2}),\theta)\\&-C(F_{1}(x_{1}),F_{2}(x_{2}-1),\theta)+C(F_{1}(x_{1}-1),F_{2}(x_{2}-1),\theta)\nonumber
	\end{align}
	The key benefit of this theory is that copulas avoid the assumption of normality when modelling non-normal data and allow the marginal distributions and the dependence structure to be estimated separately as they provide a natural way to study and measure the dependence among random variables.
	
	In this paper we used the normal copula to model the dependence between correlated binary outcomes. The normal copula \cite{meyer2013gaussian} is the most commonly used copula of the Elliptical family of copulas and can be described with the following form:
	
	\begin{equation}
		C_\Sigma^G(u_1,u_2,\rho) = \Phi_2(\Phi^{-1}(u_1),\Phi^{-1}(u_2) | \ \Sigma),
	\end{equation}
	where $\Phi_2(\cdot| \Sigma)$ is the cdf of a bivariate standard normal distribution $N(0,\Sigma)$ with covariance matrix $\Sigma$ and $\Phi^{-1}$ is the inverse cdf of the standard univariate normal distribution. The normal copula interpolates from the Frechet lower bound $\rho\rightarrow -1$ (perfect negative dependence) to the Frechet upper bound $\rho\rightarrow1$ (perfect positive dependence). 
	
	As the bivariate normal copula does not have a closed form but can be evaluated numerically using Owen's $T$-function \cite{owen1956tables}. Therefore, the normal copula can be described in terms of $T$-function with the following expression\cite{meyer2013gaussian}:
	
	\begin{align}
		\label{copula}
		C(u_1,u_2,\rho) &= \frac{u_1+u_2}{2} - T(\Phi^{-1}(u_1),a_{u_1}) - T(\Phi^{-1}(u_2),a_{u_2}) - \delta(u_1,u_2)
		\end{align}
		where
		\begin{align}
		\delta(u_1,u_2)  & = \Bigg\{  
		\begin{matrix}
			\frac{1}{2} \ \ \text{if}\ \  u_1<\frac{1}{2},\ \  u_2\ge \frac{1}{2} \text{or}\ \  u\ge \frac{1}{2}, \ \ u_2 <\frac{1}{2}  \\ 0 \ \ \text{else} 
		\end{matrix}
		\end{align}
	\text{and}\\
	\begin{align}
		a_{u_1} &= \frac{1}{\sqrt{1-\rho^2}} 	\begin{pmatrix}
			\frac{\Phi^{-1}(u_2)}{\Phi^{-1}(u_1)}-\rho \ \ 
	\end{pmatrix}, \ \	a_{u_2} = \frac{1}{\sqrt{1-\rho^2}} 
	\begin{pmatrix}
	\frac{\Phi^{-1}(u_1)}{\Phi^{-1}(u_2)}-\rho \ \ 
\end{pmatrix}	
	\end{align}
	
	\subsubsection{Bivariate random effects meta-analysis with bivariate copulas (BRMA-BC)}\label{copmodel}
	
	BRMA-IB model assumes independence of the numbers of events across arms and outcomes and accounts only for the correlation at the between-studies level. However, when modeling correlated binary outcomes (surrogate endpoint and final outcome) this assumption is too strong. As highlighted previously, at the within-study level, the numbers of events in each arm on the first and the second outcome are obtained from the same patients and are therefore correlated. Additionally, as discussed by Riley et al. \cite{riley2017multivariate}, the heterogeneity of the treatment effects on both outcomes across studies generates the between-studies correlation. Hence, two sources of association exist in the data: at the within-study level and at between-studies level.
	
	To account for the within-study association on the binomial scale, (without transforming the data to log odds ratios), the numbers of events on both outcomes should be modelled jointly, assuming association between them. This can be achieved by using a copula representation with discrete (binomial) marginals, as copulas account for the dependence between marginals and allow for modelling various dependence structures, providing a flexible representation of the bivariate distribution. Therefore, a joint density constructed with copulas can be much more flexible compared to the bivariate normal distribution which only allows for normal marginals and linear dependence structure.  
	
	At the within-study level, we assume that the summarised events in each arm on both outcomes follow bivariate distributions $h(p_{1i},p_{2i},N_{i}, \rho_{i})$ with binomial marginal distributions. The parameters $p_{1Ai}$, $p_{2Ai}$, $p_{1Bi}$, $p_{2Bi}$ denote the true probabilities of the numbers of events in each arm on the first and the second outcome, $N_{Ai}$ and $N_{Bi}$ are the numbers of patients in the control arm $A$ and experimental arm $B$ in trial $i$. 
	\begin{gather*}
	\begin{pmatrix}r_{1Ai}\\
	r_{2Ai}
	\end{pmatrix}
	\sim h(p_{1Ai},p_{2Ai},N_{Ai}, \rho_{Ai})\quad
	\begin{pmatrix}r_{1Bi}\\
	r_{2Bi}
	\end{pmatrix}	\sim h(p_{1Bi},p_{2Bi},N_{Bi}, \rho_{Bi})\\
	g(p_{1Ai})=\mu_{1i},\quad g(p_{1Bi})=\mu_{1i}+\delta_{1i}\quad
	g(p_{2Ai})=\mu_{2i},\quad g(p_{2Bi})=\mu_{2i}+\delta_{2i}\\
	\begin{pmatrix}\delta_{1i}\\
	\delta_{2i}
	\end{pmatrix} \sim  N
	\begin{pmatrix}
	\begin{pmatrix}
	d_{1}\\
	d_{2}
	\end{pmatrix}\!\!,&
	\begin{pmatrix}
	\tau_{1}^{2} & \tau_{1}\tau_{2}\rho_{b} \\
	\tau_{1}\tau_{2}\rho_{b} & \tau_{2}^{2}
	\end{pmatrix}
	\end{pmatrix}
	\end{gather*}
	where
	\begin{align*}h(r_{1Ai}, r_{2Ai} | p_{1Ai},p_{2Ai},N_{Ai}, \rho_{Ai}) &=  C(F_{1}(r_{1Ai}),F_{2}(r_{2Ai}),\rho_{Ai})-C(F_{1}(r_{1Ai}-1),F_{2}(r_{2Ai}),\rho_{Ai})\\&-C(F_{1}(r_{1Ai}),F_{2}(r_{2Ai}-1),\rho_{Ai})+C(F_{1}(r_{1Ai}-1),F_{2}(r_{2Ai}-1),\rho_{Ai}),
	\end{align*}
	\begin{align*}h(r_{1Bi}, r_{2Bi} | p_{1Bi},p_{2Bi},N_{Bi}, \rho_{Bi}) &=  C(F_{1}(r_{1Bi}),F_{2}(r_{2Bi}),\rho_{Bi})-C(F_{1}(r_{1Bi}-1),F_{2}(r_{2Bi}),\rho_{Bi})\\&-C(F_{1}(r_{1Bi}),F_{2}(r_{2Bi}-1),\rho_{Bi})+C(F_{1}(r_{1Bi}-1),F_{2}(r_{2Bi}-1),\rho_{Bi})
	\end{align*}
	$F_{1}(r_{1Ai})$, $F_{2}(r_{2Ai})$ and $F_{1}(r_{1Bi})$, $F_{2}(r_{2Bi})$ are the CDFs of the binomial marginal distributions on the two outcomes and $C(\cdot,\cdot)$ is the bivariate normal copula.
	
	Additionally, $\rho_{Ai}$, $\rho_{Bi}$ are the dependence parameters (within-study correlations) in each arm respectively and they, similarly as the within-study correlation in BRMA model, are assumed to be known. In practice, they are not reported and can be estimated using bootstrapping, as discuss in the next section, when IPD are available. 
	The within-study correlations are different across studies and hence each study has a different dependence parameter. However, in cases where IPD are not available for all studies, the same dependence parameter can be assumed across the studies, which, for example, can be an average obtained from studies with IPD available. In the absence of IPD, informative prior distributions can be constructed combining evidence from external sources such as observational studies. 
	
	At the between-studies level, the model is exactly the same as (eq. \ref{eq:4}) in BRMA-IB . The true probabilities of events $p_{1Ai}$, $p_{2Ai}$, $p_{1Bi}$, $p_{2Bi}$ are transformed using a link function $g(\cdot)$ and the true treatment effects on both outcomes are normally distributed and correlated. This model was implemented in the Bayesian framework assuming the same prior distributions as BRMA-IB. 
	
	Overall, BRMA-BC is less restrictive compared to BRMA and BRMA-IB, as it accounts for the within-study association and models the data on the original binomial scale, avoiding the a potentially inappropriate normal approximation for the marginal distributions.

	\subsection{Bootstrap methods}\label{BootMeth}
	As discussed in Section \ref{BRMA} the within-study correlations are needed to populate the correlation matrix of the BRMA model (eq. \ref{eq:1}). These correlations can be estimated when IPD are available by using a bootstrap method. The method estimates the correlation between the estimated treatment effects on both outcomes obtained by bootstrap samples with replacement from IPD \cite{daniels1997meta}. In this paper we aim to apply the method to binomial data, thus the number of events should be calculated and then transformed to log odds ratio scale for each bootstrap sample by using standard formulas. Having many pairs of the treatment effects (log ORs) on the surrogate endpoint and on the final outcome obtained from multiple bootstrap samples, it is possible not to calculate the Pearson's correlation between the treatment effect on the two outcomes.
	
	BRMA-BC model accounts for the within-study correlations on the original binomial scale by modeling the number of events on both outcomes jointly via bivariate normal copulas. Therefore, similarly to BRMA model, a bootstrap method needs to be used to populate the dependence parameters $\rho_{Ai}$ and $\rho_{Bi}$ of the model. In this case, the number of events in each arm should be calculated for each bootstrap sample and then the dependence parameter of the copula function (which defines the within-study correlation between the number of events on the first and the second outcome) can be estimated by using a optimiser such as the one available within the command $nlminb$ in R \cite{rpackage}.

	The code for the bootstrap methods can be found in the supplementary materials A.1 and A.2.

	\subsection{Implementation}\label{Implementation}

	We conducted Bayesian analysis implementing the models in cmdstanR 2.28.0 \cite{cmdstanr}.  Posterior estimates were obtained after running 4 chains consisting of 2000 MCMC iterations each after discarding 1000 iterations as warm-up period. To achieve better convergence non-centred parameterisations were used across all the models.  Convergence was visually assessed by checking rank plots, trace plots and using the diagnostic metrics of cmdstanR such as $\widehat{R}$. $\widehat{R}$ is probably the most widely used diagnostic \cite{rubin2013bayesian}. Stan uses rank-normalized folded-split $\widehat{R}$ proposed by Vehtary \cite{vehtari2021rank}. Traditionally a threshold of 1.05 is used to determine convergence, however recently Vehtari et al. suggested that a new more strict convergence threshold $\widehat{R}<$1.01 \cite{vehtari2021rank}.
	The Stan code of the models  as well as the analysis of convergence for the simulation study and the data example can be found in Sections A.3, A.4, A.5, A.9, A.10 of the supplementary materials

	The within-study association parameters for each model (BRMA and BRMA-BC) were estimated by using the two bootstrap methods (discussed in section \ref{BootMeth}). Whilst the calculation of the within-study correlation (Pearson's correlation) for BRMA model is computationally trivial via R, the maximum likelihood estimation of the dependence parameters for BRMA-BC can be computationally difficult. To estimate the dependence parameters, a two-stage estimation procedure was used. The first stage included maximum likelihood estimation of univariate binomial marginal distributions, and the second stage involved maximum likelihood estimation of the dependence parameters of the bivariate copula function with the univariate parameters from the first stage \cite{joe2005asymptotic} being held fixed. The maximum likelihood estimations were performed using $nlminb$ optimiser in $R$ \cite{rpackage}. The convergence status of the method was assessed by checking the convergence argument of the $nlminb$ function. This ensures that the optimiser provides reliable estimates.

	\section{Results of CML data example}
	
	In this section, we present results of applying the existing methodology (BRMA, BRMA-IB) and the proposed model BRMA-BC to a motivating data example in CML.
	The aim of this analysis is to evaluate the study level surrogate relationship between the candidate endpoint (CCyR) at 12 months and the final outcome (EFS) at 24 months using BRMA-BC, BRMA-IB and BRMA models. %Although BRMA and BRMA-BC models make different assumptions about the within-study variability, both of them account for the within-study association.
	As discussed in Section 3, the within-study association between treatment effects on the two outcomes can be estimated using a bootstrap method from IPD. However, in this data-set IPD were not available for any of these studies, hence we were unable to estimate the dependence parameters $\rho_{A}$ and $\rho_{B}$ of BRMA-BC and Pearson's within-study correlations $\rho_{w}$ of BRMA. Instead, we constructed informative prior distributions for each of the parameters using external evidence obtained from three observational cohort studies \cite{kantarjian2008cytogenetic,jabbour2011achievement,de2008imatinib}. These studies measured the impact of achieving a CCyR at 1 year on EFS. They reported rates of CCyR at 1 year and the rates of EFS at 2 years for the patients who either did or did not achieve CCyR at 1 year. Having this information, pseudo IPD could be generated for each of the studies, and hence the within-study associations could be estimated for each arm. Figure 1 displays the three density distributions derived from the cohort studies using double bootstrapping. The first two density distributions correspond to the prior distributions for the dependence parameters $\rho_{A}$ and $\rho_{B}$ of BRMA-BC model and the other one to the Pearson's within-study correlation $\rho_w$ used to populate BRMA.
	
	\begin{figure}[h!]
		\label{f4}
		\caption{Within-study associations}
		\centering
		\includegraphics[height= 5cm, width = 15cm]{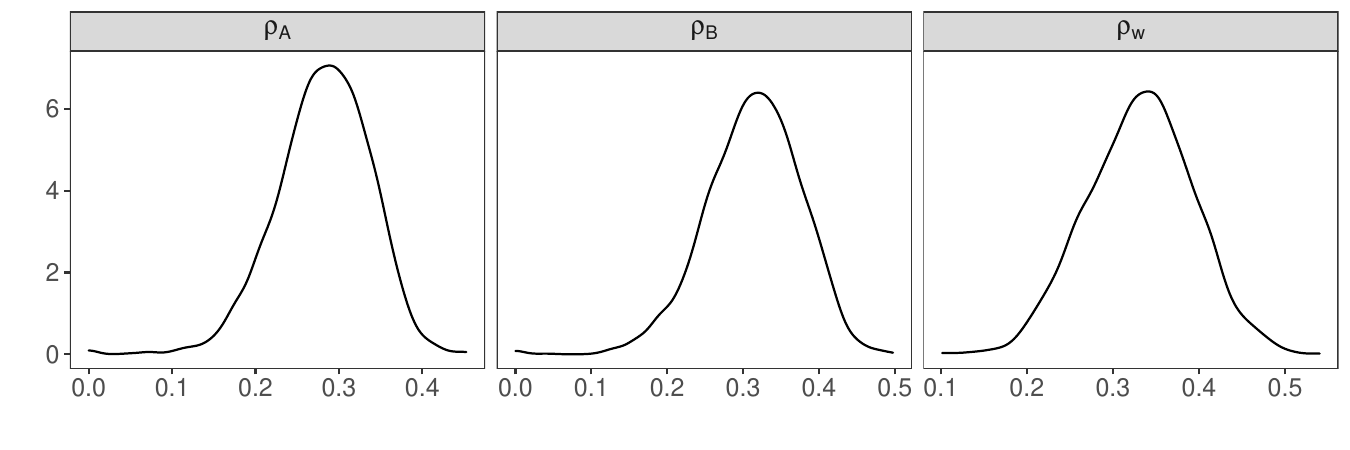}
	\end{figure}	
	
	We assumed the same prior knowledge for the within-study association parameters across all studies. Vague prior distributions were placed on all the other unknown parameters as described in Section 3. 
	
	Estimates of the between-studies parameters were obtained by running 4 chains. The convergence of the estimates was assess visually and by checking Rhat diagnostics. Detailed trace plots are presented in the supplementary material in section A.
	
	Table \ref{resultsCML} shows the estimates (means, medians and 95\% CrIs) of the between-studies parameters.
	BRMA model yielded a posterior distribution of $\rho_b$ with smallest posterior median (0.37) and the widest 95\% CrI compared to the other two models; however, all three 95\% CrIs of $\rho_b$ were very wide spanning almost from -1 to 1. 
	Similarly, 	BRMA model resulted in the smallest posterior means/medians of the heterogeneity parameters $\tau_1$, $\tau_2$ and of the pooled effects $d_1$, $d_2$, whereas their CrIs were narrower compared to the CrIs of BRMA-IB and BRMA-BC. On the other hand, BRMA-IB gaved estimates with the largest values for these parameters in these data resulting also in wider 95\% CrIs for the heterogeneity parameters and the pooled treatment effects. BRMA-BC resulted in higher posterior means and median of the between-studies parameters compared to BRMA model but slightly lower than BRMA-IB.
	
	\begin{table}[h!]
		\caption{Between-studies estimates across models}
		\label{resultsCML}
		\centering
		\begin{tabular}{lcccccccc}   
			\hline
			Models&\multicolumn{2}{c}{BRMA}&&\multicolumn{2}{c}{BRMA-BC}&&\multicolumn{2}{c}{BRMA-IB}\\
			\cline{2-3}\cline{5-6}\cline{8-9}
			Measures &Mean (Median)&95\% CrI&&Mean (Median)&95\% CrI&&Mean (Median)&95\% CrI\\
			\hline
			Parameters&&&&&&&&\\
			$\rho_{b}$&0.23(0.36)&(-0.93, 0.97)&&0.34(0.50)&(-0.89, 0.97)&&0.44(0.61)&(-0.83, 0.98)\\
			$\tau_{1}$&0.40(0.38)&( 0.11, 0.84)&&0.43(0.40)&( 0.14, 0.90)&&0.46(0.43)&( 0.16, 0.95)\\
			$\tau_{2}$&0.24(0.20)&( 0.01, 0.73)&&0.28(0.24)&( 0.02, 0.80)&&0.32(0.28)&( 0.02, 0.86)\\
			$d_{1}$   &0.47(0.46)&( 0.14, 0.81)&&0.48(0.48)&( 0.14, 0.86)&&0.49(0.49)&( 0.13, 0.88)\\
			$d_{2}$   &0.27(0.27)&(-0.04, 0.61)&&0.30(0.30)&(-0.02 0.64)&&0.30(0.30)&(-0.04, 0.67)\\	
			\hline	
		\end{tabular}
	\end{table}
	
	Overall, we drew very similar inferences about the study level surrogate relationship between the treatment effects on CCyR at 1 year and EFS at 2 years regardless of the model we used. The study level association was suboptimal as the credible intervals of the between-studies correlation were very wide, spanning almost from -1 to 1. This implies that CCyR at 12 months cannot be considered as a valid surrogate endpoint for EFS at 24 months. This is possibly due to the lack of evidence of treatment effect on EFS at 24 months, as many studies have previously reported \cite{cortes2011bosutinib,kantarjian2012dasatinib,kantarjian2011nilotinib}. 
	
	To get a better understanding of the discrepancies between estimates of the models in this data example and to illustrate the motivation and the application of the proposed method in a more detailed and controlled manner we carried out a simulation study.

	\section{Simulation study}\label{Simstdy}
	The presented methods make different assumptions at the within-study level. BRMA models the treatment effects for binomial data using normally distributed log OR scale. BRMA-IB assumes that the numbers of events across outcomes are independent and binomially distributed, whereas BRMA-BC models the numbers of events on both outcomes jointly accounting for the dependence between them at the within-study level. 
	We carried out a simulation study to assess the performance of BRMA, BRMA-IB and BRMA-BC models and in particular to investigate the impact of the assumptions made at the within-study level on estimates of the parameters at the between-studies level ($\rho_b$, $\tau_1$,$\tau_2$, $d_1$, $d_2$). 
	
	\subsection{Simulation scenarios varying within-study association, proportions of events and numbers of participants}\label{scenarios}
	
	We simulated data under 12 scenarios generating 1000 replications for each of them and varying the within-study association, the proportions of events and the numbers of participants. 	
		
	When investigating the impact of different modeling assumptions about the within-study variability on the model performance (in terms of estimating the between-studies parameters), we anticipated that such impact may depend on the strength of the within-study association. To explore this, we varied the strength of the association by assuming weak, moderate and strong within-study associations (see details in step 6 of the generation process below).
	To test the effect of the magnitude of the proportions of events on the performance of the models, we considered two sets of scenarios, one with average proportions of events equal to 0.5 and one with  high average proportions of events (0.95). This was implemented by varying the mean baseline treatment effects. In particular, baseline effects $\mu_{1i,2i}$ were drawn from the bivariate normal distribution (see details in step 3 of the data generation below). As the baseline effects are transformed on the $logit$ scale, setting the mean baseline effects $\eta_{1,2}=0$ correspond to 0.5 proportion of events in the control arm (as $logit^{-1}(0) = 0.5$), and similarly, $\eta_{1,2}=3$ correspond to 0.95 proportion of events in the first arm and on both outcomes. Lastly, we considered two settings for study sizes. The study size in both arms of each study were drawn from the following normal distribution: $n_{Ai,Bi}\sim N(m, 5)$ where $i=1,...,N$ and rounded off to the nearest integer. Setting the arm size $m=400$ and $m=150$ covers two sets of scenarios one with large study size another one with small.

	We simulated data at the individual level (zeros and ones) as within-study correlations $\rho_{wi}$ and within-study dependence parameters $\rho_{Ai}$ and $\rho_{Bi}$ for the correlation between two binary outcomes in each study $i$ are needed to populate the BRMA and BRMA-BC models and as these parameters cannot be estimated from the aggregate data. All the models were fitted to the binomial aggregate data obtained from the IPD. We used generated IPD (zeros and ones) for each study to estimate the within-study correlations and the dependence parameters for the copulas by means of bootstrapping. However, in the scenarios with high proportions of events (0.95) it is likely that some studies are generated without any non-events (zeros) both on the first and the second outcome. In such cases, the bootstrap method was unable to estimate the within-study association as the variability in the IPD is zero. We addressed this by simulating studies with at least one 'zero value' either on the first or the second outcome.
	
	Furthermore, as discussed in section \ref{Implementation} the estimation of the copula dependence parameters $\rho_{A}$ and $\rho_{B}$ can be challenging. Therefore, to ensure that the optimiser provides reliable results we monitored the number of studies where the optimiser failed to converge in each scenario. In such cases, the estimated value of $\rho_{A}$ or $\rho_{B}$ cannot be trusted as the optimiser does not provide a reliable solution. We addressed this by re-simulating IPD for these studies until the optimiser provided a reliable solution.

	The generation process is the following:
	
	\begin{enumerate}
		\item Set the number of studies to thirty ($N= 30$).
		\item Simulate the heterogeneous arm sizes $n_{i}$ of each study $i$ from the following normal distribution $(n_{i}\sim N(m,5))$ and then round them to the nearest integer.
		\item Simulate the baseline treatment effects $\mu_{1i}$, $\mu_{2i}$ from the following bivariate normal distribution $(\mu_{1i},\mu_{2i})^{T}\sim BVN \begin{pmatrix}
		\begin{pmatrix}
		\eta_1\\
		\eta_2
		\end{pmatrix}\!\!,&
		\begin{pmatrix}
		s_{1}^{2} & s_{1}s_{2}\rho \\
		s_{1}s_{2i}\rho & s_{2}^{2}
		\end{pmatrix}
		\end{pmatrix}$, with $s_{1}=s_2=0.1$ and $\rho=0.8$.
		\item Simulate the true treatment effects from $(\delta_{1i},\delta_{2i})^{T}\sim BVN \begin{pmatrix}
		\begin{pmatrix}
		d_{1}\\
		d_{2}
		\end{pmatrix}\!\!,&
		\begin{pmatrix}
		\tau_{1}^{2} & \tau_{1}\tau_{2}\rho_{b} \\
		\tau_{1}\tau_{2}\rho_{b} & \tau_{2}^{2}
		\end{pmatrix}
		\end{pmatrix}$, with $d_1 = 0.4$, $d_2 = 0.2$, $\tau_1=0.5$, $\tau_2=0.5$, $\rho_b=0.8$. 
		\item Calculate the true probabilities of events from $p_{1Ai} = logit^{-1}(\mu_{1i})$, $p_{2Ai} = logit^{-1}(\mu_{2i})$, $p_{1Bi} = logit^{-1}(\mu_{1i}+\delta_{1i})$, $p_{2Bi} = logit^{-1}(\mu_{2i}+\delta_{2i})$ in each arm across outcomes.
		\item  To simulate (weakly,  moderately and strongly) correlated binary IPD, we used a joint density with Bernoulli marginal distributions constructed with normal copula in both arms. For each set of proportions of events (0.5, 0.95) we varied the dependence parameters to reflect low, moderate and high within-study association.The true values of dependence parameters $\rho_{A}$ and $\rho_{B}$ are presented in section A.7 of the supplementary materials.
		\item Summarise the number of events in each arm and outcome by taking the sum of the binary responses. 
	\end{enumerate}	 
	This process gives us a dataset with correlated summarised events on the first and the second outcome in each arm, as well as, the original correlated binary IPD.
	
    Although the aim of the paper was to model correlated binary outcomes both at the individual and the study level, we also considered a set of scenarios with zero within-study association simulating from independent Bernoulli distributions (modifying step 6 of the generation process). This allows us to assess the robustness of the estimates obtained from BRMA-BC to different distributional assumptions. The results from the data analysis of this set of scenarios are presented and discussed briefly in section A.8 of the supplementary materials.

    \subsection{Estimands and performance measures}\label{Estimands2}
    The primary estimand of the simulation study was the parameter of the between-studies correlation $\rho_b$. The second group of estimands of the simulation study were the heterogeneity parameters $\tau_1$, $\tau_2$, the pooled effects $d_1$, $d_2$. 
    
    To evaluate the performance of the aforementioned models, in each simulation replication, we estimated the posterior median of the between-studies correlation $\widehat{\rho}_{b}$; 95\% credible interval (CrI) of $\rho_b$; coverage probability of 95\% CrIs of $\rho_b$ and then we obtained values of bias of $\widehat{\rho}_{b}$ averaged over 1000 simulation replications; and root mean squared error (RMSE) of $\rho_b$ across 1000 simulation replications. We used the same measures to evaluate the performance of the heterogeneity parameters $\tau_1$ and $\tau_2$, and the mean treatment effects $d_1$ and $d_2$.

	\subsection{Results}\label{SimResults}
	
	The results from the data analysis of the simulation study are presented in two steps. In the first step, estimates of $\rho_{wi}$, $\rho_{Ai}$ and $\rho_{Bi}$ for each study $i$ were obtained from IPD using the bootstrap methods (described in Section \ref{BootMeth}). Section \ref{Boot1} displays the findings from the first step of the data analysis, presenting the median values of $\rho_{w}$, $\rho_{A}$ and $\rho_{B}$ estimated from 3000 bootstrap samples across 30 studies and 1000 replications iterations in each scenario. 	
	In the second step of the data-analysis, we carried out the Bayesian analyses using BRMA, BRMA-IB and BRMA-BC, obtaining samples from the posterior distributions of the between-studies parameters.  
	The results of the second step are presented in Sections \ref{Rho}, \ref{Tau} and \ref{Delta} covering all the simulated scenarios.

	\subsubsection{Within-study correlations $\rho_{w}$ and within-study dependence parameters $\rho_{A}$ and $\rho_{B}$}\label{Boot1}
	
	As discussed, within-study correlations $\rho_{w}$ and within-study dependence parameters $\rho_{A}$ and $\rho_{B}$ for each study $i$ were needed to populate the BRMA and BRMA-BC models respectively. Therefore, we simulated data at the individual level to estimate them. Table \ref{Bootstrap1} and Table \ref{Bootstrap2} give details of the empirical distributions of $\rho_{w}$,and the dependence parameters of normal copula $\rho_{A}$, $\rho_{B}$ consisting of 30000 studies (30 studies $\times$ 1000 simulation replications) assuming the number of patient in each arm on average 400 and 150 respectively. We also listed results about the number of studies where the bootstrap method initially failed to provide a reliable solution i.e. the optimiser failed to converge in the section A.10 (table 7) of the supplementary material. This behaviour was very rare and occurred mainly in the scenarios where the average proportions of events were equal to 0.95 and the size of the studies small (on average 150 participants in each arm). In such cases, these studies were re-simulated until convergence was reached. This strategy resulted in fully converged estimates of the within-study association parameters.
	
		\begin{table}[h!]
		\centering
	
		\caption{Medians, 2.5\% and 97.5\% quantiles of $\rho_{w}$, $\rho_{A}$ and $\rho_{B}$ , estimated by bootstrapping simulated IPD from all the studies (30 studies) and across 1000 simulation replications, when the number of patient in each arm and was on average 400}
			\label{Bootstrap1}
		\begin{tabular}{cccc}   		
			\hline
			&& \shortstack{Average \\Proportion \\of events = 0.5} &\shortstack{Average \\Proportion \\of events = 0.95} \\
			\shortstack{Strength of\\association}&Parameter& Median 2.5\% \& 97.5\%	& Median 2.5\% \& 97.5\%\\
			\hline
			\multirow{3}{60pt}{Low within-study association}
			&$\rho_w$   & 0.14 (0.04, 0.24)&0.14 (-0.01, 0.30) \\
			&$\rho_{A}$ & 0.14 (0.03, 0.24)&0.16 (-0.01, 0.29) \\
			&$\rho_{B}$ & 0.14 (0.04, 0.24)&0.14 (-0.04, 0.36) \\
			\hline
			\multirow{3}{60pt}{Moderate within-study association}
			&$\rho_w$   & 0.40 (0.31, 0.49)&0.39 (0.15, 0.57) \\
			&$\rho_{A}$ & 0.41 (0.32, 0.50)&0.42 (0.19, 0.62) \\
			&$\rho_{B}$ & 0.40 (0.30, 0.49)&0.40 (0.10, 0.63) \\
			\hline
			\multirow{3}{60pt}{High within-study association}	
			&$\rho_w$   & 0.71 (0.64, 0.78) &0.72 (0.50, 0.85) \\
			&$\rho_{A}$ & 0.73 (0.66, 0.79) &0.76 (0.57, 0.90)\\
			&$\rho_{B}$ & 0.71 (0.60, 0.78) &0.73 (0.47, 0.90)\\
			\hline
		\end{tabular}		
		\end{table}

	\begin{table}[h!]
		\centering
			
		\caption{Medians, 2.5\% and 97.5\% quantiles of $\rho_{w}$, $\rho_{A}$ and $\rho_{B}$ , estimated by bootstrapping simulated IPD from all the studies (30 studies) and across 1000 simulation replications, when the number of patient in each arm was on average 150}
		\label{Bootstrap2}
		\begin{tabular}{cccc}    
			\hline
			&& \shortstack{Average \\Proportion \\of events = 0.5} &\shortstack{Average \\Proportion \\of events = 0.95}\\
			\shortstack{Strength of\\association}&Parameter& Median 2.5\% \& 97.5\%	& Median 2.5\% \& 97.5\%\\
			\hline
			\multirow{3}{60pt}{Low within-study association}
			&$\rho_{w}$ & 0.13 (-0.04, 0.28)&0.13 (-0.09, 0.40) \\
			&$\rho_{A}$ & 0.13 (-0.04, 0.29)&0.16 (-0.06, 0.49) \\
			&$\rho_{B}$ & 0.12 (-0.04, 0.30)&0.14 (-0.06, 0.55) \\
			\hline
			\multirow{3}{60pt}{Moderate within-study association}
			&$\rho_{w}$ & 0.40 (0.25, 0.54)&0.38 ( 0.04, 0.67)  \\
			&$\rho_{A}$ & 0.41 (0.25, 0.55)&0.43 (-0.02, 0.76) \\
			&$\rho_{B}$ & 0.40 (0.24, 0.55)&0.42 (-0.03, 0.79) \\
			\hline
			\multirow{3}{60pt}{High within-study association}	
			&$\rho_{w}$ & 0.72 (0.60, 0.80)&0.71 (0.33, 0.95)  \\
			&$\rho_{A}$ & 0.73 (0.61, 0.83)&0.77 (0.41, 0.98) \\
			&$\rho_{B}$ & 0.71 (0.57, 0.82)&0.75 (0.05, 0.98) \\
			\hline
		\end{tabular} 

		\end{table}

	\newpage
	
	\subsubsection{Between-studies correlation}\label{Rho}
	The between-studies correlation is the main parameter of interest in this paper as it quantifies the study level association between the treatment effects on the first (surrogate endpoint) and the second outcome (final outcome).
	
	Figure \ref{RhoCrIs} displays posterior medians and 95\% CrIs of $\rho_b$ averaged over the 1000 replications along with the true value of $\rho_b=0.8$ (dotted line). The plot on the left hand side (LHS), presents the results of the scenarios with large study size (the numbers of patients in both arms were simulated from $n_{Ai,Bi}\sim N(400,5)$), whereas the plot on the right hand side (RHS) illustrates the results of the scenarios with small study size (the numbers of patients in both arms were simulated from $n_{Ai,Bi}\sim N(150,5)$).
	
	Starting from the scenarios where the proportions of events were on average 0.5 and the study size was large (LHS plot, first column), BRMA and BRMA-BC models performed very similarly regardless of the strength of the within-study association. They resulted in narrow 95\% CrIs and accurate posterior medians (the average median estimate was very close to the true value). On the other hand, when the within-study association was strong, BRMA-IB model was the least accurate method overestimating between-studies correlation $\rho_b$.
	
	The next set of scenarios (LHS plot, second column) include 0.95 average proportions of events and large study size. BRMA-IB, BRMA-BC models outperformed BRMA model in terms of precision. However, BRMA-IB model was very sensitive to the effect of within-study association. The higher was the strength of the within-study association the less accurate the method was, resulting in accurate posterior medians only in the scenario with weak within-study association. Furthermore reduced uncertainty around the estimate (compared to BRMA) was observed in all scenarios but this was more pronounced in the scenarios with moderate and strong within-study association. 
	
	To investigate the effect of study size we repeated the same analysis reducing the number of participants in each study. The second plot in Figure \ref{RhoCrIs} presents the results of the scenarios with small study size (on average 150 participants in each arm). Starting from the scenarios with 0.5 average proportions of events (RHS plot, first column), BRMA, BRMA-BC were less precise but equally accurate compared to the scenarios with large study size (LHS plot, first column) resulting in very similar posterior medians, but wider 95\% CrIs. On the other hand, BRMA-IB was more susceptible to the effect of study size in terms of accuracy compared to the other two methods. Specifically, when the within-study association was either moderate or strong in scenarios presented on the RHS plot, BRMA-IB overestimated $\rho_b$ resulting in larger posterior medians compared to the corresponding scenarios of the LHS plot and the true value.
	
	The last set of scenarios (RHS plot, second column) corresponds to average proportions of events equal to 0.95 and small number of participants. In this extreme set of scenarios, all three methods performed poorly in terms of estimating $\rho_b$. This was mainly due to the extreme characteristics of this scenarion (small study size combined with the high proportions of events). Additionally, BRMA model resulted in the least accurate posterior medians and the widest 95\% CrIs. On the other hand, BRMA-IB was the most accurate and precise method.

	Figure 3 presents the bias of $\widehat{\rho}_{b}$ averaged over the 1000 replications along with the coverage probabilities of the 95\% CrIs of $\rho_b$ and RMSE of the posterior median of between-study correlation $\widehat{\rho}_b$ across the 12 scenarios. 
	It can be seen that when the average proportions of events on the first and the second outcome were 0.5 (first column, LHS, RHS plots), BRMA and BRMA-BC models performed very similarly across all three performance measures (bias, coverage, RMSE) regardless of the study size (large or small). Specifically, there was no difference in their performance across the different strengths of within-study associations, as both methods account for them. On the other hand, when within-study association was moderate (with small study size) or strong, BRMA-IB was on average the least accurate method resulting in, on average, higher biases, RMSEs and under-coverage compared to the other two methods. Concerning the effect of study sample size, in the set of scenarios with small study size, the average biases and RMSEs were substuntially higher compared to the scenarios with large study size across all methods.
	
	When the average proportions of events were 0.95 (second column, LHS and RHS plots), BRMA-BC and BRMA-IB methods outperformed BRMA model across all scenarios regardless of the study size. BRMA model substantially underestimated the between-studies correlation $\rho_{b}$ in particular when the study size was small. In the set of scenarios where the study size was large and the within-study association was moderate or strong BRMA-BC was more accurate compared to BRMA-IB resulting also in coverages closer to 95\%. On the other hand the RMSEs of BRMA-BC were higher than RMSEs of BRMA-IB. This implies that the standard error of the estimates of BRMA-BC was larger compared to those obtained form BRMA-IB despite being on average less biased across the 1000 replications (i.e. posterior medians were more dispersed around the true value).
	The estimate of the between-studies correlation of BRMA-IB was upwardly biased when the study size was large and some under-coverage was also observed when the within-study association was strong i.e., BRMA-IB produces overly optimistic 95\% CrIs of the between-studies correlation.

	\begin{landscape}
		\pagestyle{plain}
	\begin{figure}[h!]
			\centering
			\caption{Posterior medians (black dot) and 95\% CrIs (solid bars) of $\rho_b$ averaged over the 1000 replications along with the true value of $\rho_b=0.8$ (dotted line) across the 12 scenarios}
			\label{RhoCrIs}
			\includegraphics[height=14cm,width=24cm]{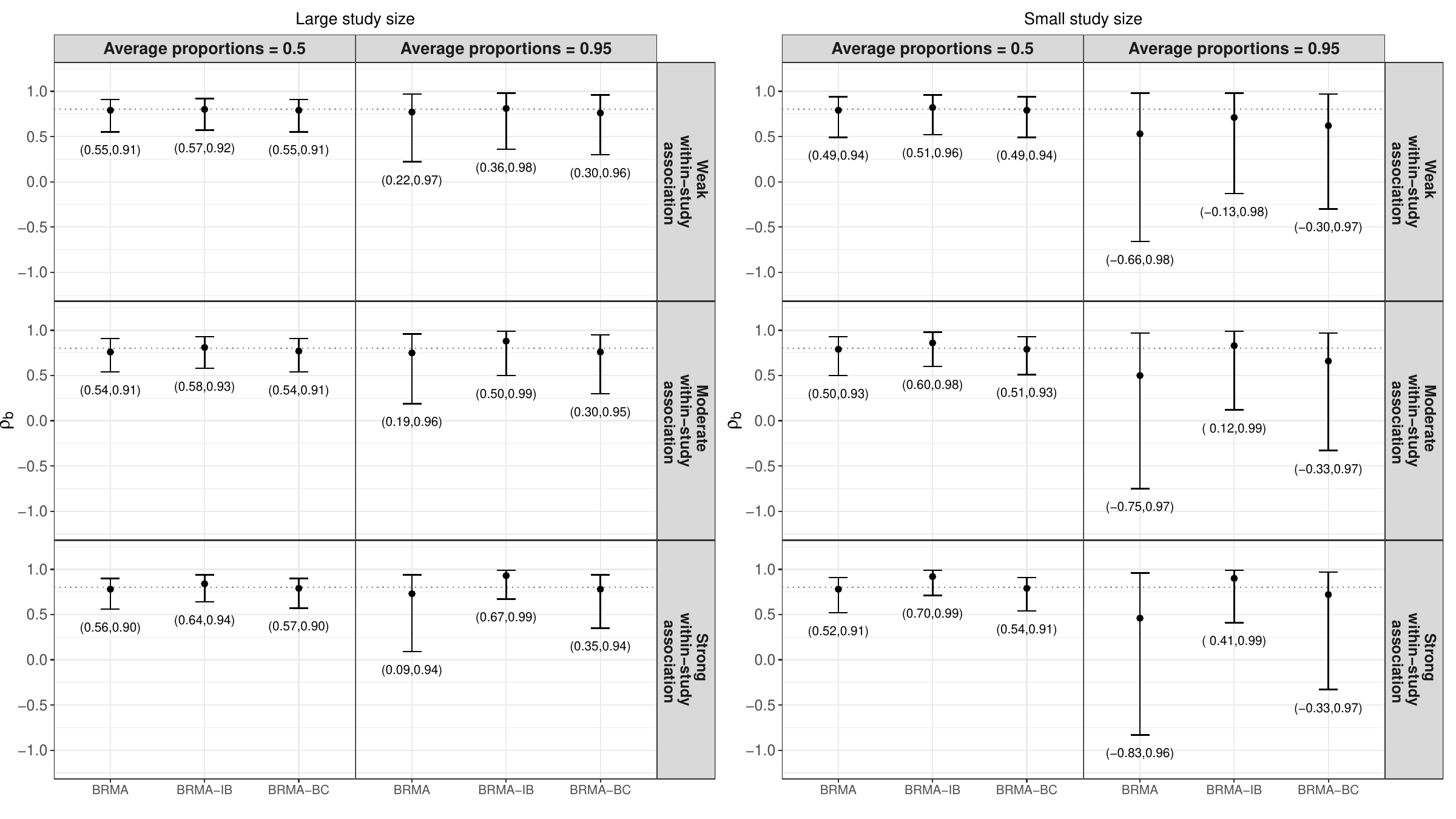}
		\end{figure}
	\end{landscape}

	\begin{landscape}
	\pagestyle{plain}	
	\begin{figure}[h!]
		\label{f1}
		\caption{Bias of $\widehat{\rho}_{b}$ averaged over the 1000 replications along with the coverage probabilities and RMSE across the 12 scenarios}
		\centering
		\includegraphics[height=13cm,width=24cm]{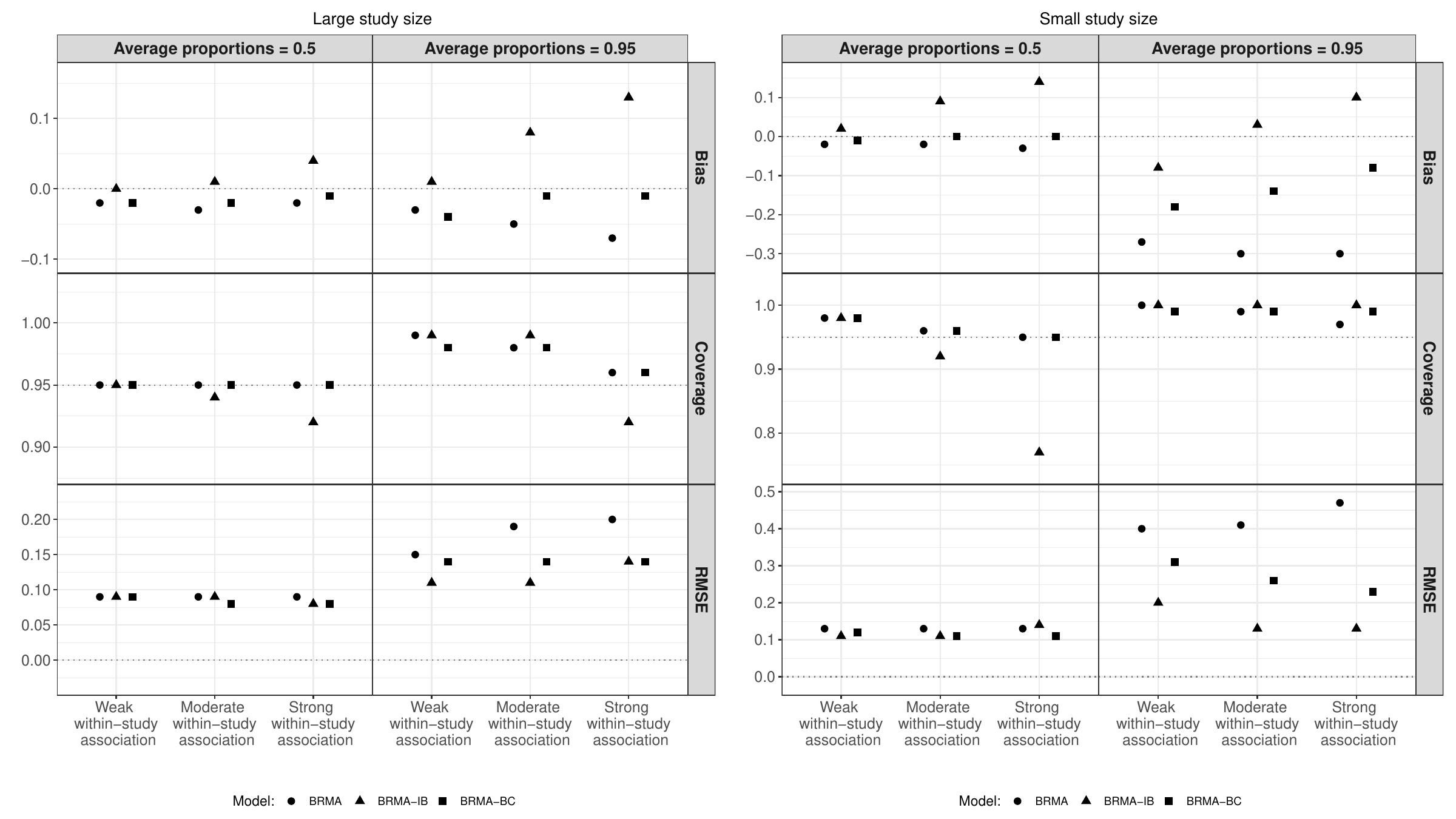}
	\end{figure}
	\end{landscape}

	\newpage
	\subsubsection{Heterogeneity parameters}\label{Tau}
	To have a better understanding of the behaviour of the between-study covariance matrix we also monitored the heterogeneity parameters (between-studies standard deviations) $\tau_{1}$, $\tau_{2}$. We report only the performance of the estimate of $\tau_{2}$ as $\widehat{\tau}_{1}$ performed in a very similar way. Figure 4 presents the bias of $\widehat{\tau}_{2}$ averaged over the 1000 replications along with the coverage probabilities of the 95\% CrIs of $\tau_2$ and RMSE across the 12 scenarios.
	
	When the average proportions of events were 0.5 (first column, LHS and RHS plots) all methods were on average unbiased, with coverage probabilities equal to 95\% and small RMSEs regardless of study size.
	
	When the average proportions of events were 0.95 (second column, LHS and RHS plots), BRMA model substantially underestimated $\tau_{2}$ across all strengths of within-study association regardless of the sample size.
	Furthermore, substantial under-coverage was observed from BRMA model when the within-study association was moderate or strong implying that BRMA resulted in overconfident 95\% CrIs.
	
	%Note that under- or overestimation of the heterogeneity parameters affects the estimates of the between-studies  correlation, which explains why $\rho_b$ obtained from BRMA was underestimated. 
	BRMA-IB overestimated the heterogeneity parameter $\tau_{2}$ mainly when the within-study association was moderate or strong and the proportions of events were 0.95. This can be associated with the upwardly biased estimates of the between-studies correlation from this method particularly in these scenarios. BRMA-BC was the most accurate method in this set of scenarios outperforming BRMA and BRMA-IB models. It yielded the most robust results achieving acceptable coverages when the study size was large, on average very accurate estimates and relatively small RMSEs. Only in the extreme scenario were the study size was small and the within-study association was strong, it resulted in downwardly biased estimates of the heterogeneity parameters.
	 
	Overall, these findings indicate that BRMA and BRMA-IB models were not appropriate methods when the proportions of events are close to 1 and the strength of the within-study association moderate or strong. 
	
	\subsubsection{Mean treatment effects }\label{Delta}
	The last set of results presents the performance of the methods in terms of the estimate of the mean treatment effect on the second (final) outcome $d_2$. The parameters of the mean effects are the main parameters of interest in the general meta-analytic framework.
	Figure 5 lists the bias of $\widehat{d}_{2}$ averaged over the 1000 replications along with the coverage probabilities and RMSE across the 12 scenarios. Similarly as in the previous section, we decided to only present results of $\widehat{d}_{2}$, as the estimates of the treatment effect on the first outcome performed in a very similar way. 
	
	When the average proportions of events were 0.5 (first column of the LHS and RHS plots) all methods performed well and in a very similar way achieving zero bias,95\% coverage probabilities and low RMSE regardless of the strength of the within-study association and the number of participants in each study. 
	
	In the second set of scenarios where the average proportions of events were 0.95 (second column of the LHS and RHS plots), BRMA gave downwardly biased estimates of $d_2$, reduced coverages and marginally higher RMSEs compared to BRMA-IB and BRMA-BC models, indicating that the assumption of normality was not reasonable. Another interesting finding was the impact of the magnitude of the within-study association on the estimates of the pooled effect $d_2$. The stronger was the within-study association more under-coverage was observed for the estimates of BRMA. This means that BRMA produced narrower 95\% CrIs than it should have been. On the other hand, BRMA-BC and BRMA-IB were less biased compared to BRMA resulting also in lower RMSEs and acceptable coverage probabilities.
	
.

	\begin{landscape}
		\pagestyle{plain}	
	\begin{figure}[h!]
		\label{f2}
		\caption{Bias of $\widehat{\tau}_{2}$ averaged over the 1000 replications along with the coverage probabilities and RMSE across the 12 scenarios}
		\centering
		\includegraphics[height=13cm,width=24cm]{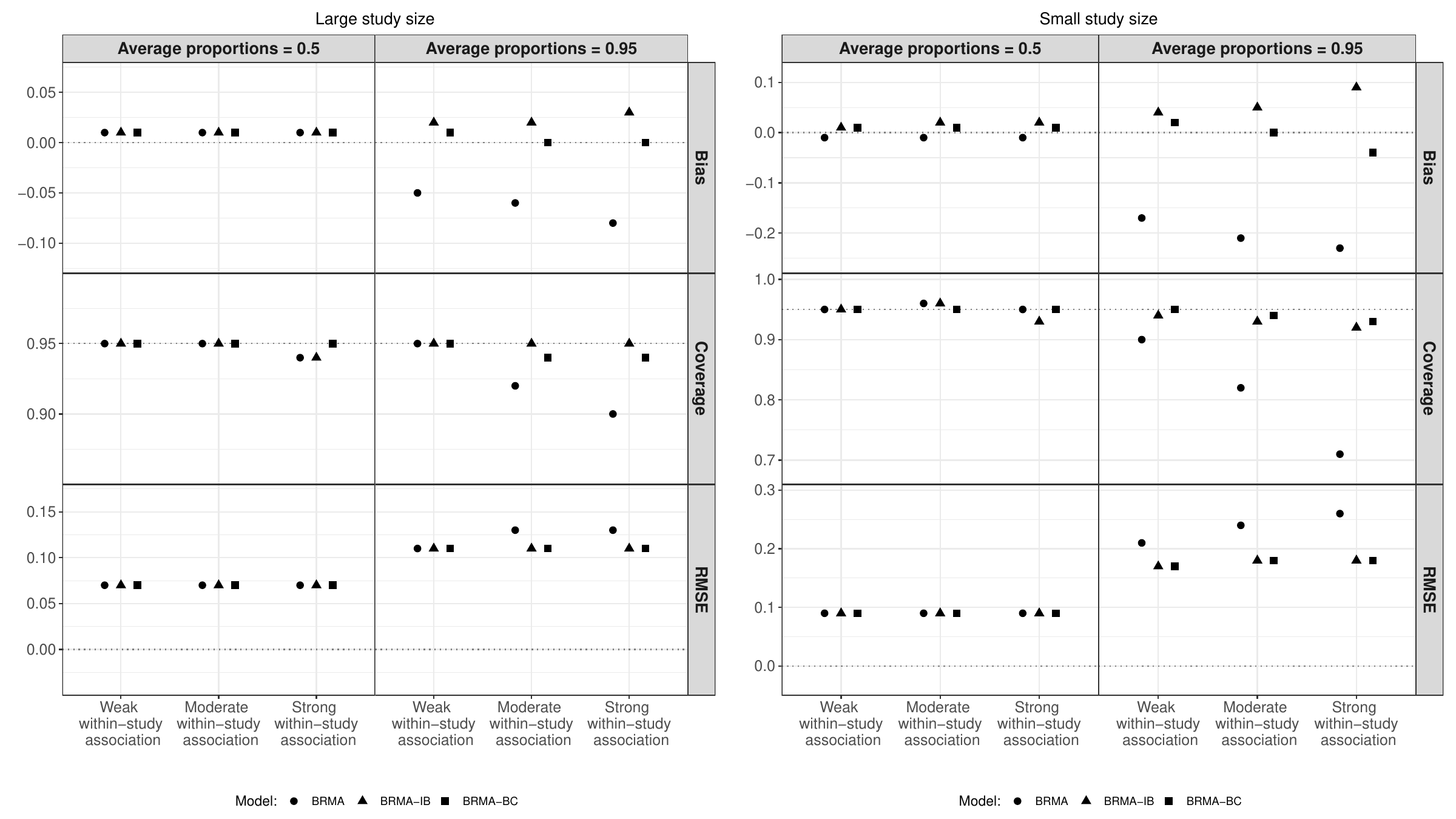}
	\end{figure}
	\end{landscape}

	\newpage

		\begin{landscape}
		\pagestyle{plain}	
	\begin{figure}[h!]
		\label{f2}
		\caption{Bias of $\widehat{d}_{2}$ averaged over the 1000 replications along with the coverage probabilities and RMSE across the 12 scenarios}
		\centering
		\includegraphics[height=13cm,width=24cm]{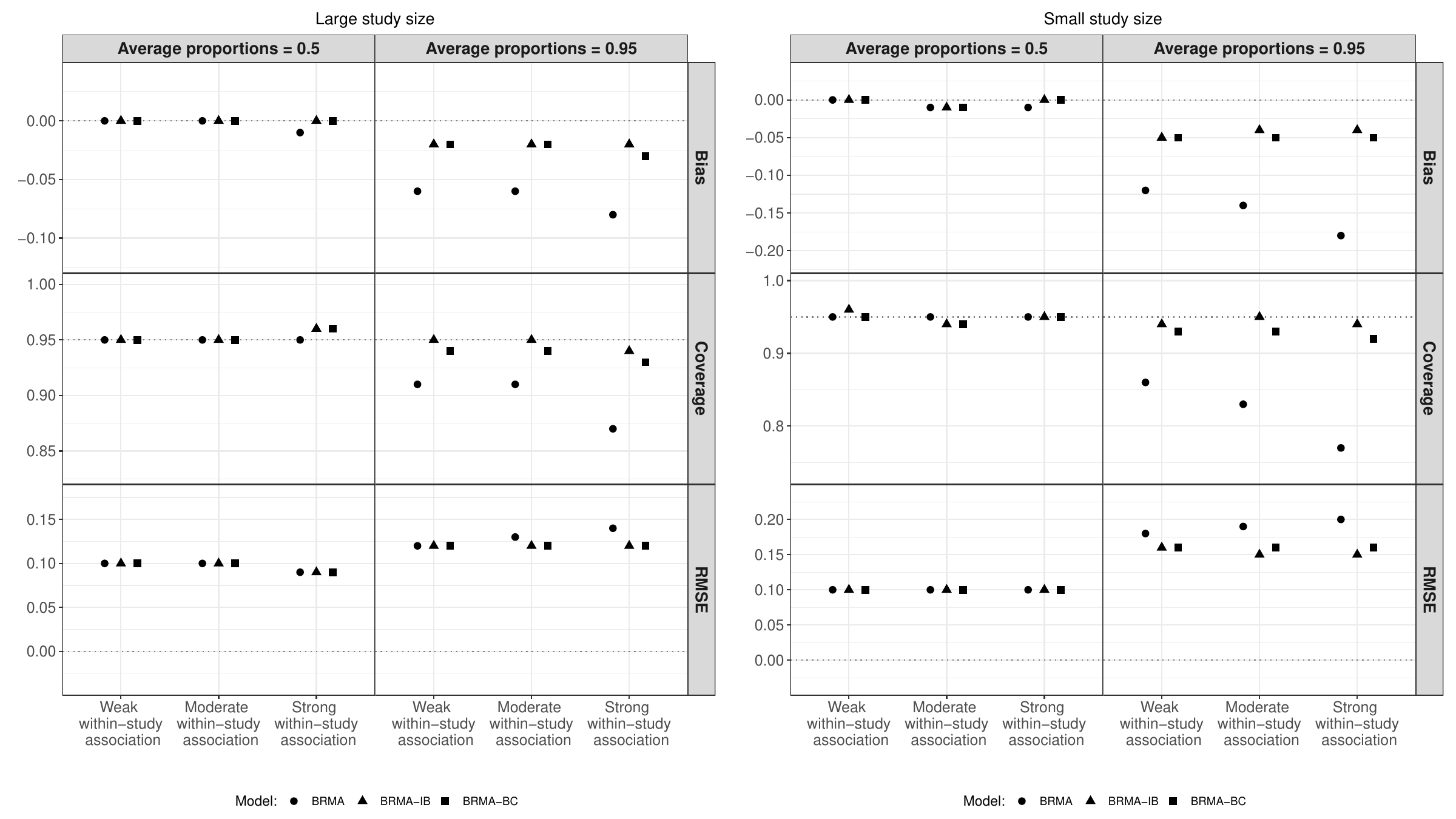}
	\end{figure}
	\end{landscape}
	\newpage

	\subsection{Key findings}
	A short summary of the key findings from the simulation study is given below:
	\begin{itemize}
		
		\item The simulation study showed that the normal approximation fails for binary outcomes when the proportions of events are close to one or zero. This confirms findings by Hamza \emph{et al}. \cite{hamza2008binomial} for the univariate case and extends their finding to the bivariate setting for binomial RCT data on two outcomes and two treatment arms. In our simulation study we focused on the performance of the parameters describing the between-studies variability: the between-studies correlation $\rho_b$ and heterogeneity parameters $\tau_{1}$, $\tau_{2}$. When the average proportions of events were 0.5, there was no clear difference between BRMA model and BRMA-BC as they performed very similarly and sufficiently well. However, when the average proportions of events were increased to 0.95. BRMA model was not appropriate to estimate the between-studies parameters as it resulted in poor coverage probabilities i.e, its 95\% were more inflated than they should have been, large RMSEs and downwardly biased estimates of $\rho_{b}$, $\tau_{1,2}$ and $d_{1,2}$.

		\item The main aim of the simulation study was to explore the impact of the within-study association on the estimation of the between-studies parameters when using the alternative modeling approaches. 
		As discussed above, BRMA model accounts for within-study association between the treatment effects on two outcomes. However, it is a suitable method for estimating the between-studies parameters only when the proportions of events are close to 0.5.
		BRMA-IB model was the most sensitive method to the effect of within-study association by far. This model assumes that the binomially distributed numbers of events are independent across outcomes. As a result, within-study associations are not taken into account and the ”excess” of the association manifests itself in the upwardly biased estimate of the between-studies correlation. In the simulation study, higher within-study associations led to more upwardly biased estimates of $\rho_b$ and substantial under-coverage.		
		BRMA-IB estimated $\rho_b$ with better precision and compared to BRMA-BC due to the fact that it overestimated the heterogeneity parameters. 
		Overall, BRMA-IB model is quite robust when modeling data with modest within-study association, but inappropriate to estimate between-studies parameters when moderate or high within-study association is present.
		
		\item The simulation study also investigated the effect of study size by having two sets of scenarios. Overall, in the scenarios with small study size, all the methods resulted in higher biases and larger RMSEs across all methods. Furthermore, the simulation study highlighted the importance of study size in the scenarios with high proportions of events. Specifically, in the scenarios with average proportions of events equal to 0.95 and small study sizes, BRMA-BC failed to estimate the trial-level association with reasonable precision despite modeling the within-study variability on the original binomial scale and accounting for within-study associations. This indicates that, the study size is important and can substantially affect the accuracy of the estimates of the between-studies correlation when investigating binary outcomes with very high/low proportions of events.
		
		\item BRMA-BC was the most appropriate method to investigate the study level association patterns between treatment effects on two binary outcomes. The model performed sufficiently well in most of the scenarios without substantially over/underestimating the heterogeneity parameters and the mean effects resulting also more accepptable coverage probabities than BRMA. There were scenarios where it failed to estimate $\rho_b$ as accurately as BRMA-IB. As explained in the previous paragraphs, this was due to the small size of the studies combined with the high proportions of events. In practice, investigating between-studies association between treatment effects on correlated binary outcomes with proportions of events close to one or zero requires studies with sufficiently large number of participants.
			
	\end{itemize}

	\section{Discussion}\label{Discussion}
	We have introduced a new bivariate meta-analytic method (BRMA-BC) and modified an existing method (BRMA-IB) which allow for modelling the within-study variability of the binomial data on the original binomial scale. BRMA-BC offers a robust framework for meta-analysis of binary outcomes avoiding the use of an unreliable approximation of normality for log odds ratios. In this paper, we used the proposed methodology to improve the evaluation of study level surrogate relationships between treatment effects  at both the within-study and the between-study level on two binary outcomes with high proportions of events.
	This can be particularly useful in diseases where the increased effectiveness of targeted treatments often leads to high numbers of responses and reduced numbers of events.
		
	Standard meta-analytic methods, such as BRMA, can model the observed treatment effects using a bivariate normal distribution of log odds ratios. Although this approach accounts for the within-study association, the assumption of normality for the marginal distributions is unreasonable when the proportions of events are close to 1, leading to biased results. BRMA-IB model avoids the assumption of normality but it is more restrictive compared to BRMA-BC. It models the within-study variability using binomial likelihoods; however, it ignores the within-study association. BRMA-BC models the data on the original binomal scale and accounts for both sources of association (within-study and between-studies correlation). At the within-study level, it models the numbers of events on the surrogate endpoint and the final outcome jointly using bivariate distributions constructed with copulas. The model is very flexible as it can also account for different dependence structures between the marginal distributions by using different types of copulas. 
	
	BRMA-IB model performs well when the within-study association is weak regardless of the size of the studies. In such scenarios, it can offer substantial gains in accuracy of the estimates of the parameters describing the surrogate relationship (in particular when the proportions of events are close to one or zero), resulting also in acceptable coverage probabilities and smaller RMSEs compared to BRMA model.
	However, as the strength of the within-study association increases, the performance of the model becomes problematic. BRMA-IB ignores the within-study association and the ”excess” of the association manifests itself in the upwardly biased estimate of the between-studies correlation. For example, in the scenarios where the within-study association was moderate or strong the model failed to estimate well the between-studies variability, giving upwardly biased estimates and low coverage probabilities of the between-studies correlation $\rho_b$.
	
	BRMA-BC is the most robust model to quantify the study level association regardless of the strength of the within-study associations. In particular in the scenarios with average proportions of events of 0.95, the model resulted in less biased estimates of the between-studies correlation compared to BRMA. Furthermore, the fact that in the majority of the scenarios the model did not over/underestimate the heterogeneity parameters $\tau_{1,2}$ led to more reasonable estimates of the between-studies correlation $\rho_b$ compared to BRMA-IB. However, there were some extreme scenarios where BRMA-BC failed to yield as accurate estimates of the between-studies correlation, as well as BRMA-IB model. For instance, when the average proportions of events was 0.95 and the study size small, BRMA-BC model yielded on average downwardly biased estimates of $\rho_b$. This was due to small sample and the very small number of non-events which made the estimation of $\rho_b$ extremely difficult. Therefore, when the proportions of events are very high, a sufficiently large sample size is needed to estimate the between-studies correlation accurately. Overall, across the 12 scenarios of the simulation study, BRMA-BC model was superior to BRMA model.
	
	In the data example, all methods found suboptimal study level association between the treatment effects on CCyR at 12 months and the treatment effects on EFS at 24 months. Across methods, the posterior medians of the between-studies correlations were not very high and the corresponding 95\% CrIs were extremely wide, spanning almost from -1 to 1. However, BRMA-BC model resulted in larger between-studies correlation $\rho_b$ and slightly larger heterogeneity parameters on the first and the second outcome compared to BRMA model. Overall, the posterior medians of the between-studies parameters from BRMA model were lower compared to the other two models. This behaviour is similar to the findings of the simulation study in the scenarios with proportions of events close to 1, where BRMA resulted in the lowest estimates of the between-studies correlation and the heterogeneity parameters and the pooled effects, whilst BRMA-IB the highest. Furthermore, BRMA resulted in 95\% CrIs of the heterogeneity parameters and the pooled effects with reduced uncertainty compared to the other methods which is in line with the findings of the simulation study where, BRMA produced narrower 95\% CrIs than BRMA-BC and BRMA-IB (under-coverage was observed in many scenarios for $\tau_2$ and $d_2$). 
	
	Although BRMA-BC model provides robust results in a variety of scenarios, potential limitations should always be kept in mind. First, in order to perform Bayesian inference, we run MCMC with a No-U-turn sampler (NUTS) using cmdstanR. A limitation of the method was the fact that BRMA-BC model was very sensitive to initial values. Therefore, the initiation of the estimation process was difficult without setting "sensible" initial values. This was tackled by fitting BRMA-IB prior to BRMA-BC and then converting the estimates of BRMA-IB to initial values for BRMA-BC. However, this issue makes the use of BRMA-BC model quite restrictive, as it requires another method to be fitted prior to BRMA-BC model.
	
	A limitation of the illustrative example was the lack of IPD. We informed the prior distributions of within-study association parameters using 3 cohort studies. We constructed binary pseudo IPD and hence calculated the within-study association between the numbers of responses on the surrogate endpoint and the numbers of events on the final outcome by using a double bootstrap method to account for uncertainty. Furthermore, the definition of EFS varied across these studies with some studies presenting it as PFS and some others included a broader range of events in their definition than others. However, a sensitivity analysis showed that by excluding a small number studies where EFS was defined slightly differently did not affect the results and the inferences.  
	
	BRMA-BC can be extended in a number of ways. For instance, it can be extended by using also a copula at the between-studies level in a similar way as in Nikolopoulos et al. \cite{nikoloulopoulos2015mixed}. This will allow to model the study level association on the true scale (proportions of events) with beta marginal distributions avoiding the logit transformation. Furthermore, taking advantage of the setting proposed by Bujkiewicz et al. \cite{bujkiewicz2015bayesian}, BRMA-BC can be extended to allow for modelling multiple surrogate endpoints (or the same surrogate endpoint but reported at multiple time points) via a vine-copula. 
	In this work, the main aim was to assess the impact of modeling the binomial data on the original scale. Therefore, we used the bivariate normal copula as dependence structure to simulate data in the simulation study and to model the within-study level of BRMA-BC model. This makes the comparison between BRMA and BRMA-BC fair as the two models share the same dependence structure (linear and symmetric) at the within-study level. However, BRMA-BC is much more flexible as it can easily be implemented using alternative copulas with different dependence structures. 	
	
	In summary, we developed a new Bayesian hierarchical meta-analytic method and modified an existing method to perform bivariate meta-analysis of binary outcomes and particularly, to quantify the study level surrogate relationship. In our view, BRMA-BC is a preferred model for modelling binary outcomes in the context of surrogate endpoints, as well as, in the general meta-analytic context of multiple outcomes. The model can improve the process of the validation of surrogate endpoints in the era of personalised medicine where the increased effectiveness of targeted treatments often leads to high numbers of responses and reduced numbers of events.
	
	\section{Acknowledgements}
	
	This research used the ALICE/SPECTRE High Performance Computing Facility at the University of Leicester. The authors thank Enti Spata for sharing data from the literature review on CML. Sylwia Bujkiewicz and Keith Abrams were funded by the Medical Research Council, methodology research grant MR/T025166/1.
	
   \bibliography{bibli}{}
\end{document}